\documentclass[prl,aps,amsmath,amssymb,twocolumn,floatfix]{revtex4}

\usepackage{graphicx}
\newcommand{\vlk}{$V_{{\rm low}-k}$ }
\newcommand{\vlkn}{$V_{{\rm low}-k}$}
\newcommand{\be}{\begin{equation}}
\newcommand{\ee}{\end{equation}}

\begin{document}
\title{Shell model description of the $^{14}$C dating $\beta$ decay \\
with Brown-Rho-scaled NN interactions}

\author{J.\ W.\ Holt$^1$, G.\ E.\ Brown$^1$, T.\ T.\ S.\ Kuo$^1$,
  J.\ D.\ Holt$^2$, R.\ Machleidt$^3$}
\affiliation{$^1$Department of Physics, SUNY, Stony Brook, New York 11794,
  USA\\ $^2$TRIUMF, 4004 Wesbrook Mall, Vancouver, BC, Canada, V6T 2A3 \\ 
$^3$Department of Physics, University of Idaho, Moscow, Idaho 83844, USA}

\date{\today}

\begin{abstract}
We present shell model calculations for the $\beta$-decay of $^{14}$C
to the $^{14}$N ground state, treating the states of
the $A=14$ multiplet as two $0p$ holes in an $^{16}$O core. We employ
low-momentum nucleon-nucleon (NN) interactions derived from the realistic
Bonn-B potential and find that the Gamow-Teller (GT) matrix element is too
large to describe the known lifetime. By using a modified version of this
potential that incorporates the effects of Brown-Rho scaling medium
modifications, we find that the GT matrix element vanishes for a nuclear
density around 85\% that of nuclear matter. We find that the splitting between
the $(J^\pi,T)=(1^+,0)$ and $(J^\pi,T)=(0^+,1)$ states in $^{14}$N is improved
using the medium-modified Bonn-B potential and that the transition strengths
from excited states of $^{14}$C to the $^{14}$N ground state are compatible
with recent experiments.
\end{abstract}

\maketitle

The beta decay of $^{14}$C to the $^{14}$N ground state has long been recognized as a unique problem in nuclear structure. Its connection to the radiocarbon dating method, which has had a significant impact across many areas of science, makes the decay of broad interest even beyond nuclear physics. But {\it a priori} one would not expect the beta decay of $^{14}$C to be a good transition for radiocarbon dating over archaeological times, because the quantum numbers of the initial state $(J^\pi,T) = (0^+,1)$ and final state $(J^\pi,T) = (1^+,0)$ satisfy the selection rules for an allowed Gamow-Teller transition. The expected half-life would therefore be on the order of hours, far from the unusually long value of 5730 years \cite{ajzen} observed in nature. The corresponding nuclear transition matrix element is very small ($\simeq 2 \times 10^{-3}$) and is expected to result from an accidental cancellation among the different components contributing to the transition amplitude. This decay has therefore been used to investigate phenomena not normally considered in studies of allowed transitions, such as meson exchange currents \cite{goulard,huffman}, relativistic
effects \cite{jin}, and configuration mixing \cite{abrown,towner}. Of broader importance, however, is that this decay provides a very sensitive test for the in-medium nuclear interaction and in particular for the current efforts to extend the microscopic description of the nuclear force beyond that of a static two-body potential fit to the experimental data on two-nucleon systems. One such approach is to include hadronic medium modifications, in which the masses of mesons and nucleons are altered at finite density due to the partial restoration of chiral symmetry \cite{bernard,hatsuda,klimt} or many-body interactions with either intermediate nucleon-antinucleon excitations \cite{bwbs} or resonance-hole excitations \cite{klingl}. These effects are traditionally incorporated in models of the three-nucleon force (3NF), which have been well-tested in {\it ab initio} nuclear structure calculations of light nuclei \cite{navratil,pieper}.

In this Letter we suggest that a large part of the observed $^{14}$C beta decay suppression arises from in-medium modifications to the nuclear interaction. We study the problem from the perspective of Brown-Rho scaling (BRS) \cite{brown1,brown2}, which was the first model to make a comprehensive prediction for the masses of hadrons at finite density. In BRS the masses of nucleons and most light mesons (except the pion whose mass is protected by its Goldstone boson nature) decrease at finite density as the ratio of the in-medium to free-space pion decay constant:
\begin{equation}
\sqrt{\frac{g_A}{g_A^*}}\frac{m_N^*}{m_N} = \frac{m_\sigma^*}{m_\sigma} =
\frac{m_\rho^*}{m_\rho} = 
\frac{m_\omega^*}{m_\omega} = \frac{f_\pi^*}{f_\pi} = \Phi(n),
\label{brs}
\end{equation}
where $g_A$ is the axial coupling constant, $\Phi$ is a function of the nuclear density $n$ with $\Phi(n_0) \simeq 0.8$ at nuclear matter density, and the star
indicates in-medium values of the given quantities. Since all realistic models of the NN interaction are based on meson exchange and fit to only free-space data, eq.\ (\ref{brs}) prescribes how to construct a density-dependent nuclear interaction that accounts for hadronic medium modifications. This program has been carried out in several previous studies
of symmetric nuclear matter \cite{rapp,jeremy}, where it was found that one
could well describe saturation and several bulk equilibrium properties of
nuclear matter using such Brown-Rho-scaled NN interactions.

The case of the $^{14}$C beta decay provides a nearly ideal situation in
nuclear structure physics for testing the hypothesis of Brown-Rho scaling. Just
below a double shell closure, the valence nucleons of $^{14}$C inhabit a region with a large nuclear density. But more important is the sensitivity of this GT matrix element to the nuclear tensor force, which as articulated by Zamick and collaborators \cite{zamick,fayache} is one of the few instances in nuclear structure where the role of the tensor force is clearly revealed. In fact, with a residual interaction consisting of only central and spin-orbit forces it is not possible to achieve a vanishing matrix element in a pure $p^{-2}$ configuration \cite{inglis}. Jancovici and Talmi \cite{talmi} showed that by including a strong tensor force one could construct an interaction which reproduces the lifetime of $^{14}$C as well as the magnetic moment and electric quadrupole moment of $^{14}$N, although agreement with the known spectroscopic data was unsatisfactory.

\begin{figure}[hbt]
\includegraphics[height=8.5cm,angle=-90]{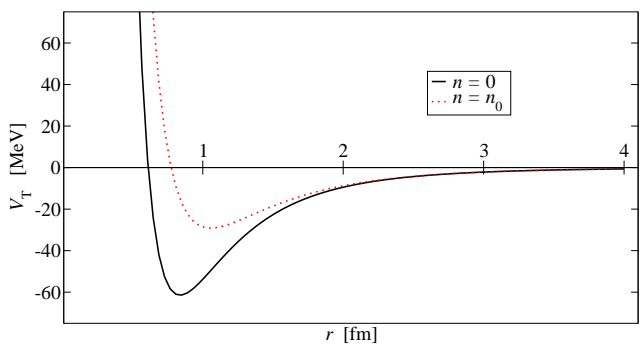}
\caption{The radial part of the nuclear tensor force given in eq.\
  (\ref{tenforce}) from $\pi$ and $\rho$ meson exchange at zero density and
  nuclear matter density under the assumption of BRS.}
\label{tensor}
\end{figure}

The most important contributions to the tensor force 
come from $\pi$ and $\rho$ meson exchange, which act opposite to each other:
\begin{eqnarray}
V_\rho^T(r) &=& \frac{f_{N\rho}^2}{4\pi}m_\rho \tau_1 \cdot \tau_2 \left(
  -S_{12} \left[ \frac{1}{(m_\rho r)^3} + \frac{1}{(m_\rho r)^2}
\right. \right. \nonumber \\
&+& \left. \left. \frac{1}{3 m_\rho r} \right] e^{-m_\rho r}\right), \nonumber
\\
V_\pi^T(r) &=& \frac{f_{N\pi}^2}{4\pi}m_\pi \tau_1 \cdot \tau_2\left(
  S_{12} 
  \left[ \frac{1}{(m_\pi r)^3} + \frac{1}{(m_\pi
      r)^2}\right. \right. \nonumber \\ 
&+& \left. \left. \frac{1}{3 m_\pi r} \right]e^{-m_\pi r} \right).
\label{tenforce}
\end{eqnarray}
Since the $\rho$ meson mass is expected to decrease substantially at nuclear
matter density while the $\pi$ mass remains relatively constant, an
unambiguous prediction of BRS is the decreasing of the tensor force at
finite density, which should be clearly seen in the GT matrix element. In
fact, recent shell model calculations \cite{fay} performed in a
larger model space consisting of $p^{-2} + 2\hbar \omega$ excitations have
shown that the $\beta$-decay suppression requires the in-medium tensor force
to be weaker and the in-medium spin-orbit force to be stronger in comparison
to a typical $G$-matrix calculation starting with a realistic NN
interaction. We show in Fig.\ \ref{tensor} the radial part of the tensor
interaction $V^T(r) = V_\pi^T(r) + V_\rho^T(r)$ at zero density and nuclear
matter density assuming that $m_\rho^*(n_0)/m_\rho = 0.80$.

Experiments to determine the properties of hadrons in medium have
been performed for all of the light mesons important in nuclear structure
physics. Studies of deeply-bound pionic atoms \cite{geissel} find only a small
increase in the $\pi^-$ mass at nuclear matter density and a related decrease
in the $\pi^+$ mass. Experimental information on the scalar and vector
particles comes from mass distribution measurements of in-medium decay
processes. Recent photoproduction experiments \cite{mess} of correlated pions
in the $T=J=0$ channel ($\sigma$ meson) have found that the distribution is
shifted to lower masses in medium. The vector mesons have been the most
widely studied. Whereas the situation is clear with the $\omega$ meson, the
mass of which drops by $\sim$ 14\% at nuclear matter density \cite{trnka},
with the $\rho$ meson it is still unclear \cite{naruki,nasseripour}. We
believe that our present study tests the decrease in $\rho$ mass more simply.

Today there are a number of high precision NN interactions based solely on
one-boson exchange. In the present work we use the Bonn-B potential
\cite{machleidt} which includes the exchange of the $\pi$, $\eta$, $\sigma$,
$a_0$, $\rho$, and $\omega$ mesons. In \cite{rapp} the consequences of
BRS on the free-space NN interaction were incorporated into the Bonn-B
potential and shown to reproduce the saturation properties of nuclear
matter in a Dirac-Brueckner-Hartree-Fock calculation. The masses of the
pseudoscalar mesons were unchanged, and the vector meson masses as well as the
corresponding form factor cutoffs were decreased according to
\begin{equation}
\frac{m_\rho^*}{m_\rho}=\frac{m_\omega^*}{m_\omega}=\frac{\Lambda^*}{\Lambda}
= 1-0.15\frac{n}{n_0}.
\end{equation}
The medium-modified (MM) Bonn-B potential is unique in its microscopic
treatment of the scalar $\sigma$ particle as correlated 2$\pi$ exchange. Finite density effects arise through medium modifications to the exchanged $\rho$ mesons in the pionic $s$-wave interaction as well as through the dressing of the in-medium pion propagator with $\Delta$-hole excitations. These modifications to the vector meson masses and pion propagator would traditionally be included in the chiral three-nucleon contact interaction and the 3NF due to intermediate $\Delta$ states, respectively.

Using realistic NN interactions in many-body perturbation theory is problematic due to the strong short distance repulsion in relative $S$ states. The modern solution is to
integrate out the high momentum components of the interaction in such a way
that the low energy physics is preserved. The details for constructing such
a low momentum interaction, \vlkn, are described in \cite{bogner02,bogner03}.
We define $V_{\rm low-k}$ through the $T$-matrix equivalence $T(p',p,p^2) =
T_{\rm low-k}(p',p,p^2)$ for $( p',p) \leq \Lambda$, where $T$ is given by the
full-space equation $T=V_{NN}+V_{NN}gT$ and $T_{\rm low-k}$ by the
model-space (momenta  $\leq \Lambda$) equation $T_{\rm low-k}=V_{\rm
  low-k}+V_{\rm low-k}gT_{\rm low-k}$.
Here $V_{NN}$ represents the Bonn-B NN potential and $\Lambda$ is the 
decimation momentum beyond which the high-momentum components of $V_{NN}$ 
are integrated out. Since pion production
starts around $E_{\rm lab} \simeq 300$ MeV, the concept of a real NN potential
is not valid beyond that energy. Consequently, we choose $\Lambda \approx 2.0$
fm$^{-1}$ thereby retaining only the information from a given potential that
is constrained by experiment. In fact for this $\Lambda$, the $V_{\rm low-k}$
derived from various NN potentials are all nearly identical \cite{bogner03}.

We use the folded diagram formalism to reduce the full-space nuclear many-body
problem $H\Psi _n=E_n \Psi _n$ to a model space problem $H_{\rm eff}\chi
_m=E_m \chi _m$ as detailed in \cite{ko90}. Here $H=H_0+V$, $H_{\rm
  eff}=H_0+V_{\rm eff}$, $E_n=E_n(A=14)-E_0(A=16,\rm core)$, and $V$ denotes
the bare NN interaction. The effective interaction $V_{\rm eff}$ is derived
following closely the folded-diagram method detailed in \cite{jensen95}. A
main difference is that in the present work the irreducible vertex function
($\hat{Q}$-box) is calculated from the low-momentum interaction \vlkn, while in
\cite{jensen95} from the Brueckner reaction matrix ($G$-matrix). In the
$\hat{Q}$-box we include hole-hole irreducible diagrams of first- and
second-order in \vlkn. Previous studies \cite{bogfurn,jason,cora07} have found that \vlk is suitable for perturbative calculations; in all of these references satisfactory converged results were obtained including terms only up to second order in \vlkn.

Our calculation was carried out in $jj$-coupling where in the basis $\left\{
    p_{3/2}^{-2},p_{3/2}^{-1} p_{1/2}^{-1}, p_{1/2}^{-2}\right\}$ one must
    diagonalize 
\be
\left[ V_{\rm eff}^{ij} \right] + \left[\begin{array}{ccc}
0 & 0 & 0 \\
0 & \epsilon & 0 \\
0 & 0 & 2\epsilon \end{array} \right],
\ee
to obtain the ground state of $^{14}$N (and a similar $2\times
    2$ matrix for $^{14}$C). We used $\epsilon = E(p_{1/2}^{-1}) -
    E(p_{3/2}^{-1}) = 6.3$ MeV, which is the experimental excitation energy of
    the first 
    $\frac{3}{2}^-$ state in $^{15}$N. One can transform the wavefunctions to
    $LS$-coupling, where the $^{14}$C and $^{14}$N ground states are
\begin{eqnarray}
\psi_i &=& x \left |^1S_0\right > + y\left|^3P_0\right >\nonumber \\
\psi_f &=& a \left |^3S_1\right > + b\left|^1P_1\right > + c\left|^3D_1\right>
\label{ls}
\end{eqnarray}
and the Gamow-Teller matrix element $M_{\rm GT}$ is given by \cite{talmi}
\begin{equation}
\sum_{k} \left<\psi_f||\sigma(k)\tau_+(k)||\psi_i\right> =
-\sqrt 6\left (xa-yb/\sqrt 3 \right).
\end{equation}
Since $x$ and $y$ are expected to have the same sign \cite{inglis}, the GT
matrix element can vanish only if $a$ and $b$ have the same sign, which
requires that the $\left<^3S_1\right|V_{\rm eff}\left|^3D_1\right>$ matrix
element furnished by the tensor 
force be large enough \cite{talmi}. In Table \ref{lstab} we show the
ground state wavefunctions of $^{14}$C and $^{14}$N, as
well as the GT matrix element, calculated with the MM Bonn-B
interaction.
\setlength{\tabcolsep}{.075in}
\begin{table}[htb]
\begin{tabular}{|c|c|c|c|c|c|c|} \hline
$n/n_0$ & $x$ & $y$ & $a$ & $b$ & $c$ & $M_{\rm GT}$ \\ \hline
0    & 0.844 & 0.537 & 0.359 & 0.168 & 0.918 & -0.615 \\ \hline
0.25 & 0.825 & 0.564 & 0.286 & 0.196 & 0.938 & -0.422 \\ \hline
0.5  & 0.801 & 0.599 & 0.215 & 0.224 & 0.951 & -0.233 \\ \hline
0.75 & 0.771 & 0.637 & 0.154 & 0.250 & 0.956 & -0.065 \\ \hline
1.0  & 0.737 & 0.675 & 0.103 & 0.273 & 0.956 &  0.074 \\ \hline
\end{tabular}
\caption{The coefficients of the LS-coupled wavefunctions defined in eq.\
  (\ref{ls}) and the associated GT matrix element as a function of the nuclear
  density $n$.}
\label{lstab}
\end{table}

\begin{figure}[hbt]
\includegraphics[height=8.5cm, angle=-90]{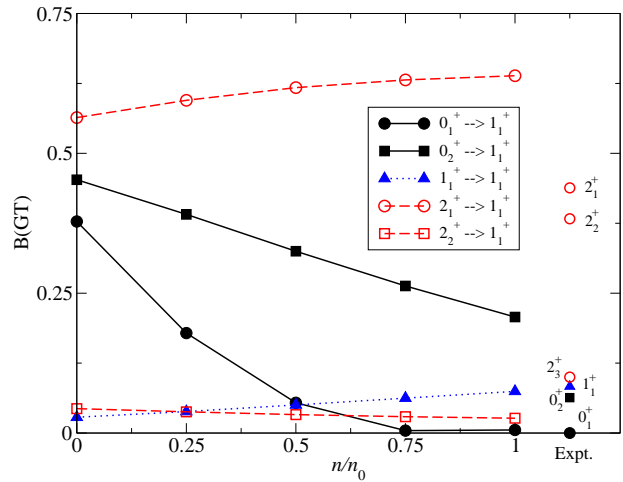}
\caption{The $B(GT)$ values for transitions from the states of $^{14}$C to the
  $^{14}$N ground state as a function of the nuclear density and the
  experimental values from \cite{negret}. Note that there are three
  experimental low lying $2^+$ states compared to two theoretical $2^+$ states
  in the $p^{-2}$ configuration.}
\label{bgt}
\end{figure}

\begin{figure}[hbt]
\includegraphics[height=8.5cm, angle=-90]{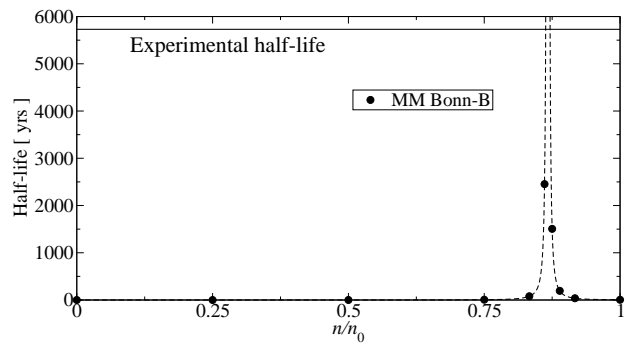}
\caption{The half-life of $^{14}$C, as a function of the nuclear density, calculated from the MM Bonn-B potential.}
\label{c14dating}
\end{figure}

In Fig.\ \ref{bgt} we plot the resulting $B(GT)=\frac{1}{2J_i+1}|M_{\rm GT}|^2$
values for transitions between the low-lying states of $^{14}$C and the
$^{14}$N ground state for the in-medium Bonn-B NN interaction
taken at several different densities. Recent experiments \cite{negret} have
determined the GT strengths from the $^{14}$N ground state to excited states
of $^{14}$C and $^{14}$O using the charge exchange reactions
$^{14}{\rm N}(d, {^{2}{\rm He}})^{14}{\rm C}$ and $^{14}{\rm N}(^3{\rm
  He},t)^{14}{\rm O}$, and our theoretical calculations are in good overall
agreement. The most prominent effect we find is a robust inhibition of the
ground state to ground state transition for densities in the range of
$0.75-1.0n_0$. In contrast, the other transition strengths are more mildly
influenced by the density dependence in BRS. In Fig.\ \ref{c14dating} we show the resulting half-life of $^{14}$C calculated from the MM Bonn-B potential.

\begin{figure}[hbt]
\includegraphics[height=8.5cm,angle=-90]{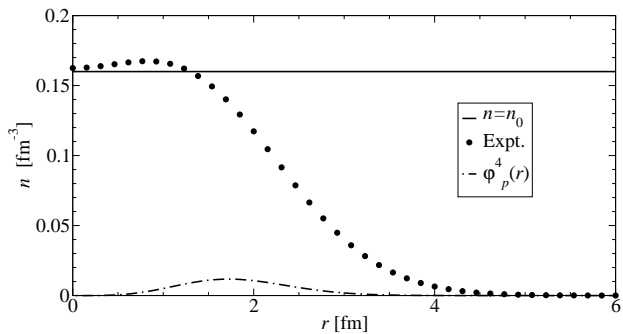}
\caption{Twice the charge distribution of $^{14}$N taken from \cite{schaller,
    schutz} and the fourth power of the $p$-shell wavefunctions.}
\label{hosc}
\end{figure}

\begin{figure}[hbt]
\includegraphics[height=8.5cm, angle=-90]{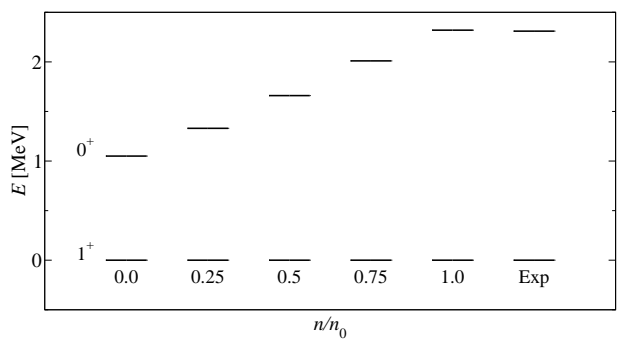}
\caption{The splitting between the $1^+_1$ and $0^+_1$ levels in $^{14}$N for
  different values of the nuclear density. Also included is the experimental
  value.}
\label{lvlsplit}
\end{figure}

We emphasize that the nuclear density experienced by $p$-shell nucleons is actually close to that of nuclear matter; in Fig.\ \ref{hosc} we compare twice the charge distribution of $^{14}$N obtained from electron scattering experiments \cite{schaller,schutz} with the radial part of the $0p$ wavefunctions, indicating clearly that the nuclear density for $0p$ nucleons is $\sim 0.8n_0$.
The first excited $0^+$ state of $^{14}$N together with the ground states of
$^{14}$O and $^{14}$C form an isospin triplet. We have calculated the splitting in energy between this state and the ground state of $^{14}$N for a range of nuclear densities. Our results are presented in Fig.\ \ref{lvlsplit},
where the experimental value is 2.31 MeV.

In summary, we have shown that by incorporating hadronic medium modifications into the Bonn-B potential the decay of $^{14}$C is strongly suppressed at densities close to that experienced by valence nucleons in $^{14}$C. In a more traditional approach such medium modifications would be built in through 3N forces, and we suggest that calculations with free-space 2N interactions supplemented with 3N forces should also inhibit the GT transition.

\begin{acknowledgments}
We thank Igal Talmi for helpful correspondences. This work was partially
supported by the U.\ S.\ Department of Energy under Grant No.\
DE-FG02-88ER40388 and the U.\ S.\ National Science Foundation under Grant No.\
PHY-0099444. TRIUMF receives federal funding via a contribution agreement through the National Research Council of Canada, and support from the Natural Sciences and Engineering Research Council of Canada is gratefully acknowledged.
\end{acknowledgments}



\end{document}